\documentclass{moriond}
\usepackage{xspace}
\usepackage{lineno}
\newcommand{\PQt}{\ensuremath{\mathrm{t}\xspace}}
\newcommand{\PQq}{\ensuremath{\mathrm{q}\xspace}}
\newcommand{\PZ}{\ensuremath{\mathrm{Z}\xspace}}
\newcommand{\PH}{\ensuremath{\mathrm{H}\xspace}}
\newcommand{\PW}{\ensuremath{\mathrm{W}\xspace}}
\newcommand{\PAQt}{\ensuremath{\bar{\mathrm{t}}}\xspace}
\newcommand{\ttbar}{\ensuremath{\PQt\PAQt}\xspace}
\newcommand{\tttt}{\ensuremath{\PQt\PAQt\PQt\PAQt}\xspace}
\newcommand{\ttt}{\ensuremath{\PQt\PAQt\PQt}\xspace}
\newcommand{\ttZ}{\ensuremath{\PQt\PAQt\PZ}\xspace}
\newcommand{\ttW}{\ensuremath{\PQt\PAQt\PW}\xspace}
\newcommand{\ttH}{\ensuremath{\PQt\PAQt\PH}\xspace}
\newcommand{\tWZ}{\ensuremath{\PQt\PW\PZ}\xspace}
\newcommand{\ttg}{\ensuremath{\PQt\PAQt\gamma}\xspace}
\newcommand{\ttgg}{\ensuremath{\PQt\PAQt\gamma\gamma}\xspace}
\newcommand{\tqg}{\ensuremath{\PQt\PQq\gamma}\xspace}
\newcommand{\pt}{\ensuremath{p_{\mathrm{T}}}\xspace}

\newcommand{\Photo}{\includegraphics[height=35mm]{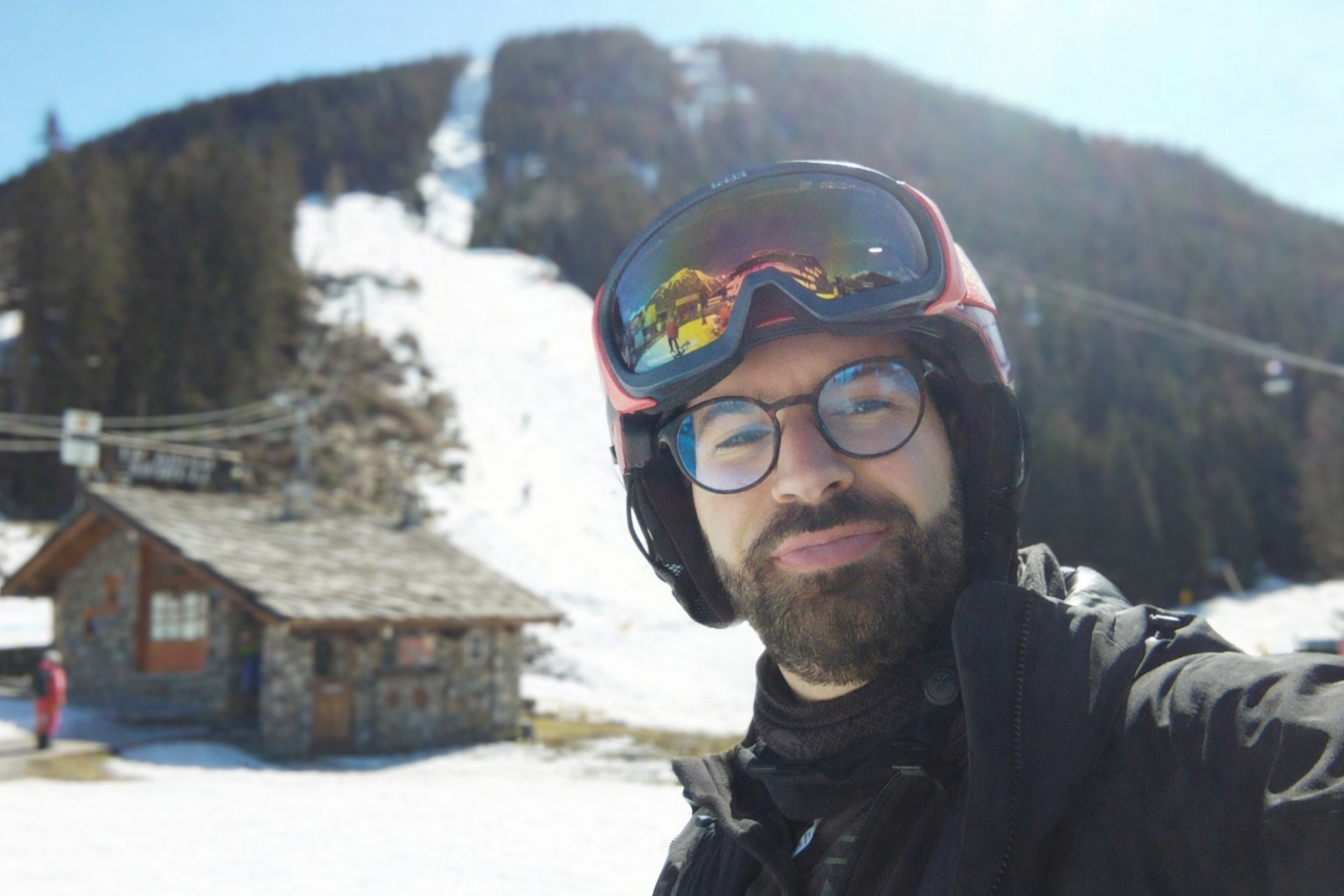}}
\begin{document}
\vspace*{4cm}
\title{Rare top quark production and top quark properties in ATLAS and CMS}

\author{ Sergio S\'anchez Cruz (on behalf of the ATLAS and CMS Collaborations)\footnote{Copyright 2024 CERN for the benefit of the ATLAS and CMS Collaborations. Reproduction of this article or parts of it is allowed as specified in the CC-BY-4.0 license.} }

\address{Universidad de Oviedo \\ C/ San Francisco, 3, 33003 \\ Oviedo, Spain}

\maketitle\abstracts{
The production of top quark pairs is one of the most relevant production modes at the LHC, and
allow for precise measurement of the properties of this particle.
Top quarks are also produced through rarer mechanisms,
including the production of multiple top quarks or the associated production of top
quarks with electroweak gauge bosons. Although these processes have significantly
smaller cross sections, they provide unique sensitivity to the couplings of the top
quark and to possible effects of physics beyond the standard model (SM). This contribution
reviews recent analyses of rare top quark production performed by the ATLAS and CMS
Collaborations.
}

\section{Introduction}

Top quarks can be produced at the LHC through a variety of production modes, with
\ttbar production providing the dominant contribution. In addition, several rare
production modes exist, typically characterized by cross sections below approximately
1~pb. These include the production of multiple top quarks, such as \tttt and \ttt,
as well as the associated production of top quarks with electroweak gauge bosons,
such as \ttZ, \ttW, \ttg, and other processes. Despite their small rates, these
processes are particularly sensitive to the structure of the top quark couplings to
the electroweak sector and therefore provide powerful probes of physics beyond the
SM. This contribution summarizes recent results from the ATLAS~\cite{atlas}
and CMS~\cite{cms1,cms2} Collaborations.

\section{Electroweak \texorpdfstring{\ttW}{ttW} production}

\begin{figure}[h]
  \centering
  $\vcenter{\hbox{\includegraphics[width=0.5\textwidth]{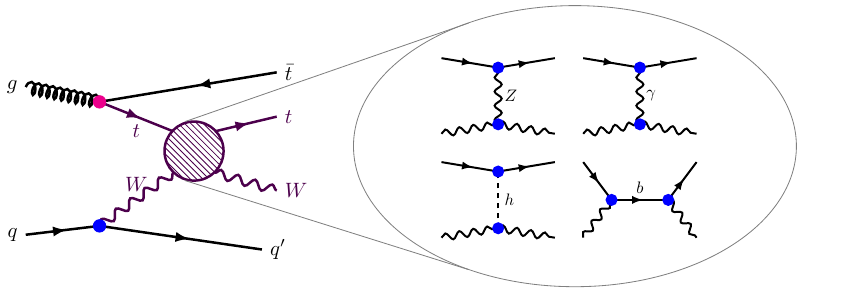}}}$

  $\vcenter{\hbox{\includegraphics[width=0.3\textwidth]{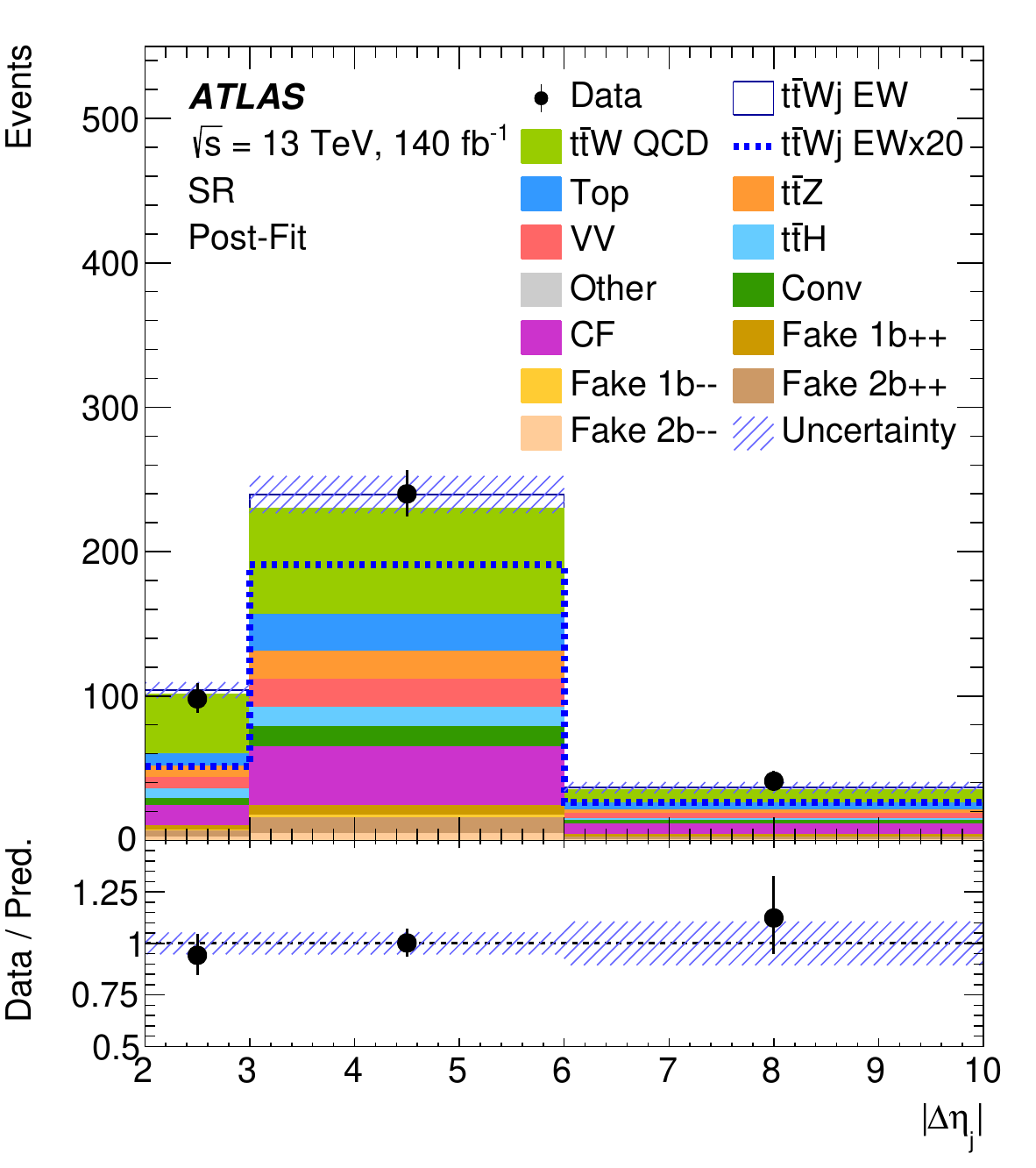}}}$
  $\vcenter{\hbox{\includegraphics[width=0.6\textwidth]{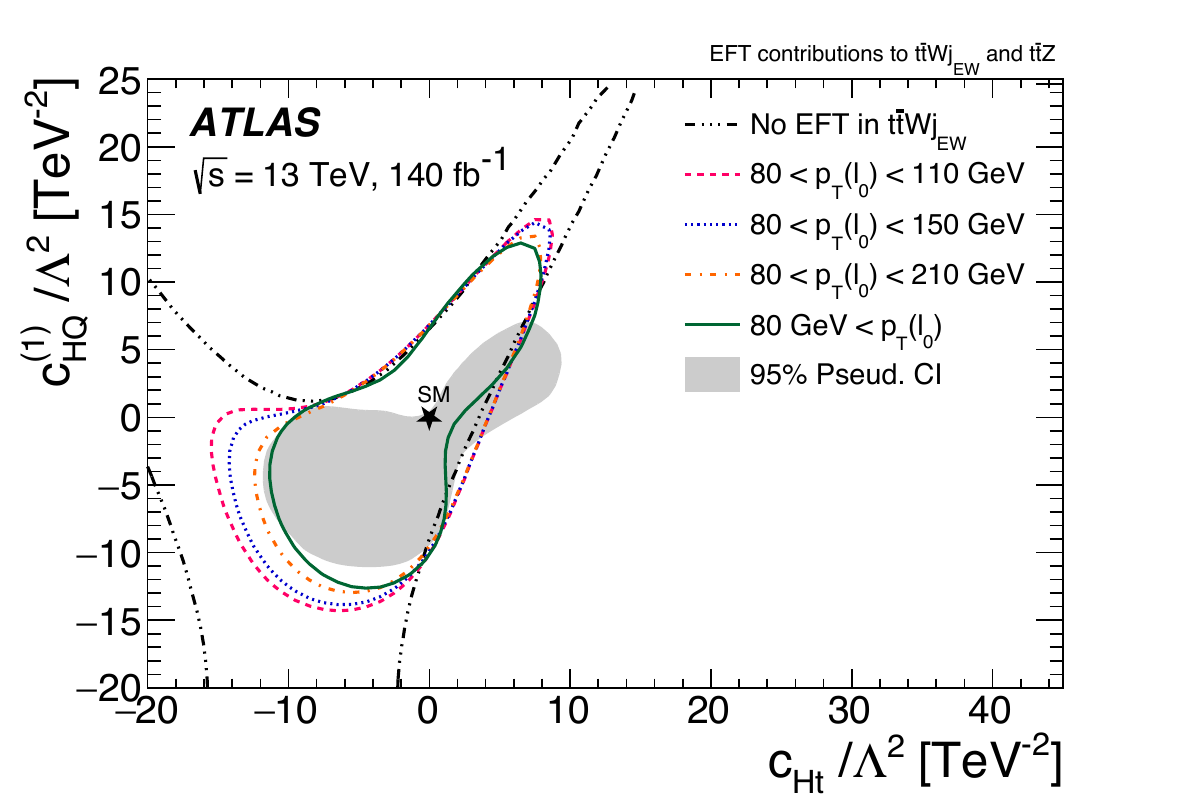}}}$
  \caption{Upper: Example of Feynman diagrams contributing to \ttW production through $\PQt\PW\to\PQt\PW$
  scattering. Lower left: $|\Delta\eta_j|$ distribution for events entering the analysis selection. Lower right: Limits
  on two effective field theory operators. All plots are taken from Ref.~\protect\cite{ttwewk}.}
  \label{fig:ttw_ewk}
\end{figure}

The modeling of \ttW production is challenging due to the sizeable contribution from
higher-order perturbative corrections, arising from additional production modes that
open at next-to-leading order and beyond. A recent result from the ATLAS
Collaboration~\cite{ttwewk} targets mixed electroweak and quantum chromodynamics (QCD)
corrections, including diagrams such as those shown in Fig.~\ref{fig:ttw_ewk}, featuring
$\PQt\PW\to\PQt\PW$ scattering. Such processes are particularly sensitive to possible
new physics effects.

The analysis uses events containing a pair of same-sign leptons (electrons or muons)
accompanied by additional hadronic activity, including a light-flavor jet that probes the
spectator quark present in the diagram. Such a selection is dominated by \ttW production,
events with non-prompt leptons or where the lepton charge has been mismeasured,
and, to a lesser extent, \ttZ and \ttH production.
Figure~\ref{fig:ttw_ewk} shows the distribution of events as a function of
the difference in pseudorapidity between the most forward
non $b$-tagged jet and the jet with which it forms the largest invariant mass, $|\Delta\eta_j|$. A fit to this distribution together with a set of control
regions designed to constrain the main backgrounds is performed
to set limits on the contribution to the
$\PQt\PW\to\PQt\PW$ scattering signal.
The analysis sets an upper limit of 251~fb at 95\%
confidence level (CL), approximately five times the SM prediction. The
analysis is also interpreted in the framework of effective field theory, allowing
constraints to be placed on operators affecting the $\PQt\PW$ interaction, as shown in
Fig.~\ref{fig:ttw_ewk} (lower right). The results are consistent with SM expectations and
highlight the complementarity between analyses using \ttZ and \ttW production. While
measurements based solely on \ttZ are insensitive to certain linear combinations of
operators, these degeneracies can be resolved when information from \ttW production is
included.

\section{Search for top quark pairs in association with leptons}

A search for the production of top quark pairs in association with additional leptons
has been reported by the ATLAS Collaboration~\cite{ttll}. This analysis is sensitive to
four-fermion operators that induce reactions where two top quarks are produced in association
with a lepton pair. 
A previous CMS measurement probed the same operators in similar topologies~\cite{top22006}.
The search selects events with significant jet activity and three leptons, including a
pair of opposite-sign same-flavor (OSSF) leptons with an invariant mass above the $\PZ$ boson
mass. The dominant contributions to this topology are \ttW production and \ttZ production,
where the $\PZ$ boson is produced off-shell. Dedicated control regions are defined
to predict and validate the description of these backgrounds.  

\begin{figure}[h]
  \centering
  \includegraphics[width=0.3\textwidth]{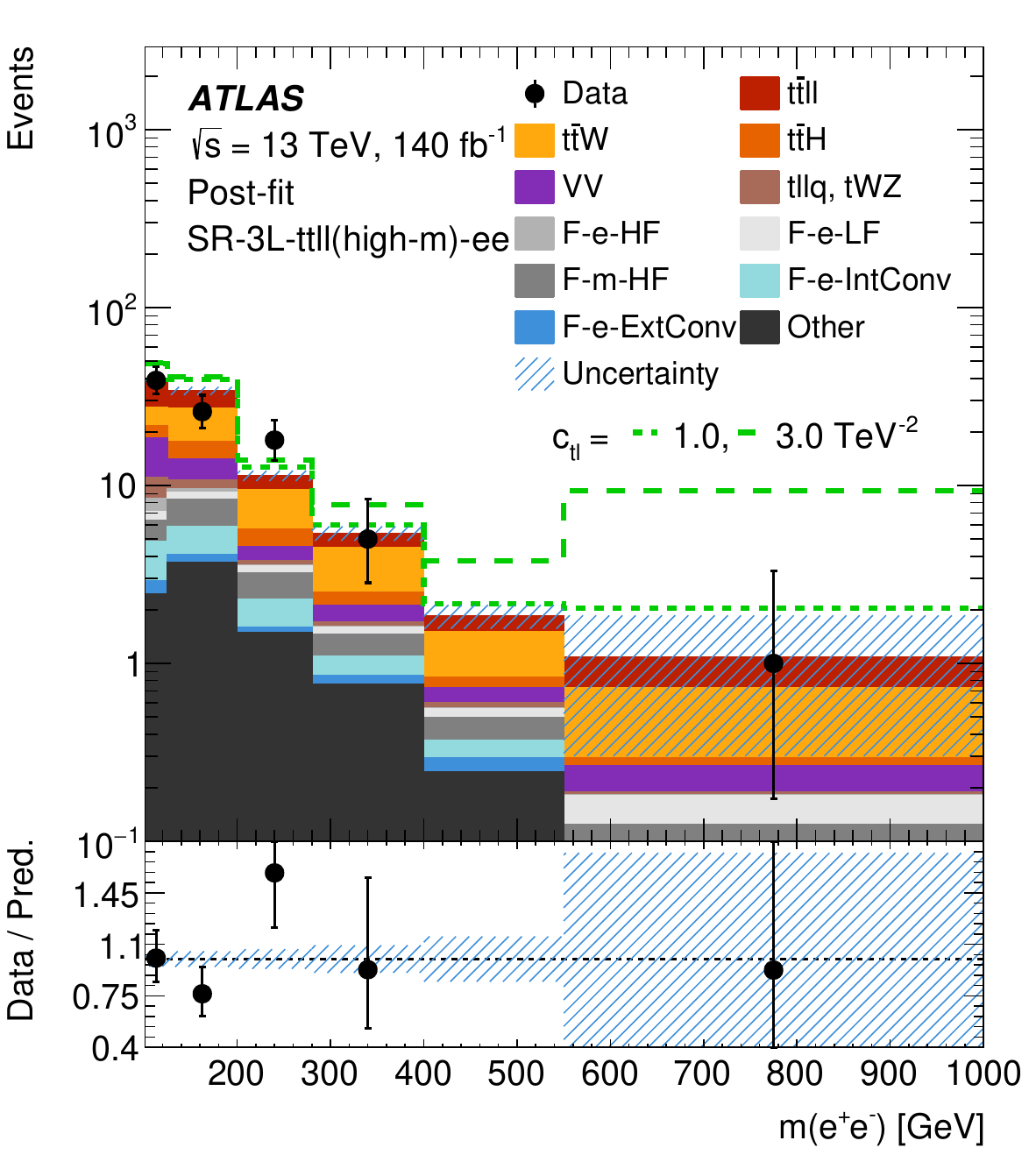}
  \includegraphics[width=0.3\textwidth]{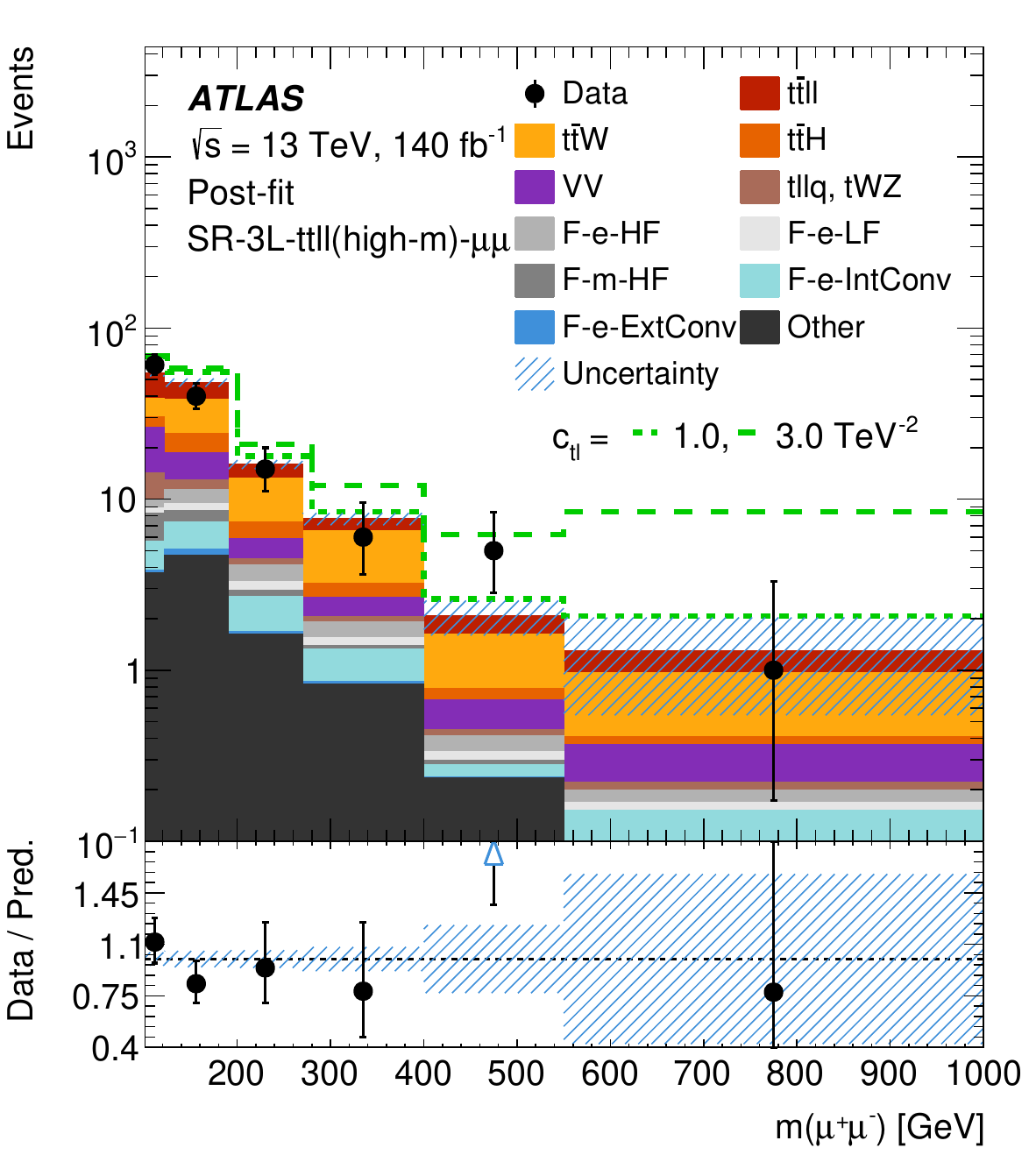}
  \includegraphics[width=0.38\textwidth]{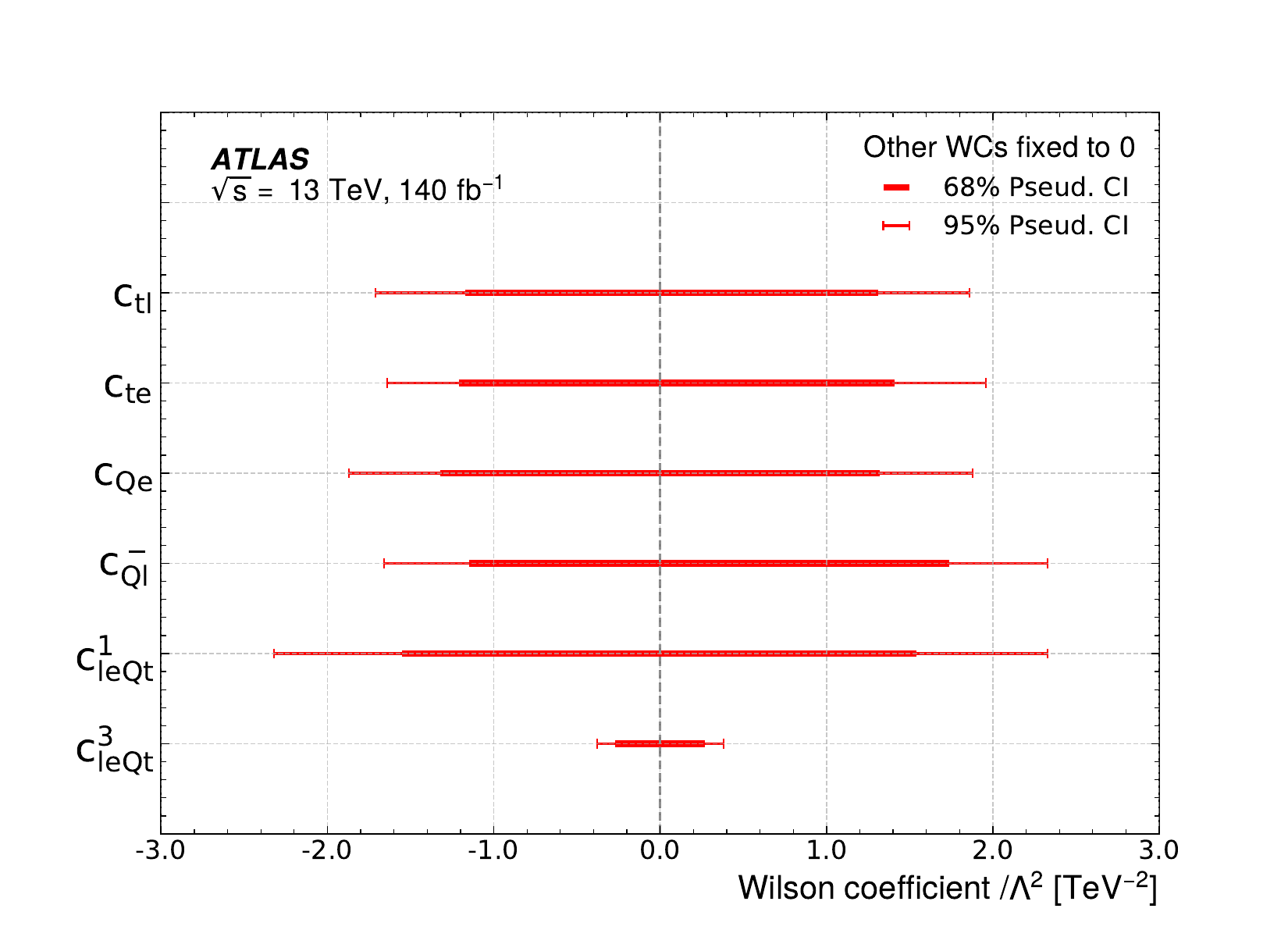}
  
  \caption{Distribution of the invariant mass of the opposite-sign same-flavor lepton pair closest
    to the $\PZ$ boson mass, separately for electrons (left) and muons (middle). Right: limits on a set of 4-fermion operators that couple top quark pairs to lepton pairs. All plots are taken from Ref~\protect\cite{ttll}.}
  \label{fig:ttll}
\end{figure}

The invariant mass distribution of the OSSF lepton pair closest to the $\PZ$ boson mass is
shown in Fig.~\ref{fig:ttll}, separately for electron and muon channels. Good agreement
is observed between the data and the SM prediction. The results are used to
set limits on Wilson coefficients parameterizing contact interactions between top quark
pairs and lepton pairs, which are shown in Fig.~\ref{fig:ttll}, for the case in which the
new physics effects contribute both to electrons and muons. The contribution also considers
scenarios in which new physics particles couple exclusively to a single lepton generation, setting separate
limits for electrons and muons. In addition, constraints are placed on the difference between the new physics
contributions to electrons and muons, thereby enhancing sensitivity to scenarios in which the new
physics does not respect lepton flavor universality.

\section{Measurement of \texorpdfstring{\ttg}{ttgamma} and \texorpdfstring{\tqg}{tqgamma} production}

\begin{figure}[h]
  \centering
  \includegraphics[width=0.7\textwidth]{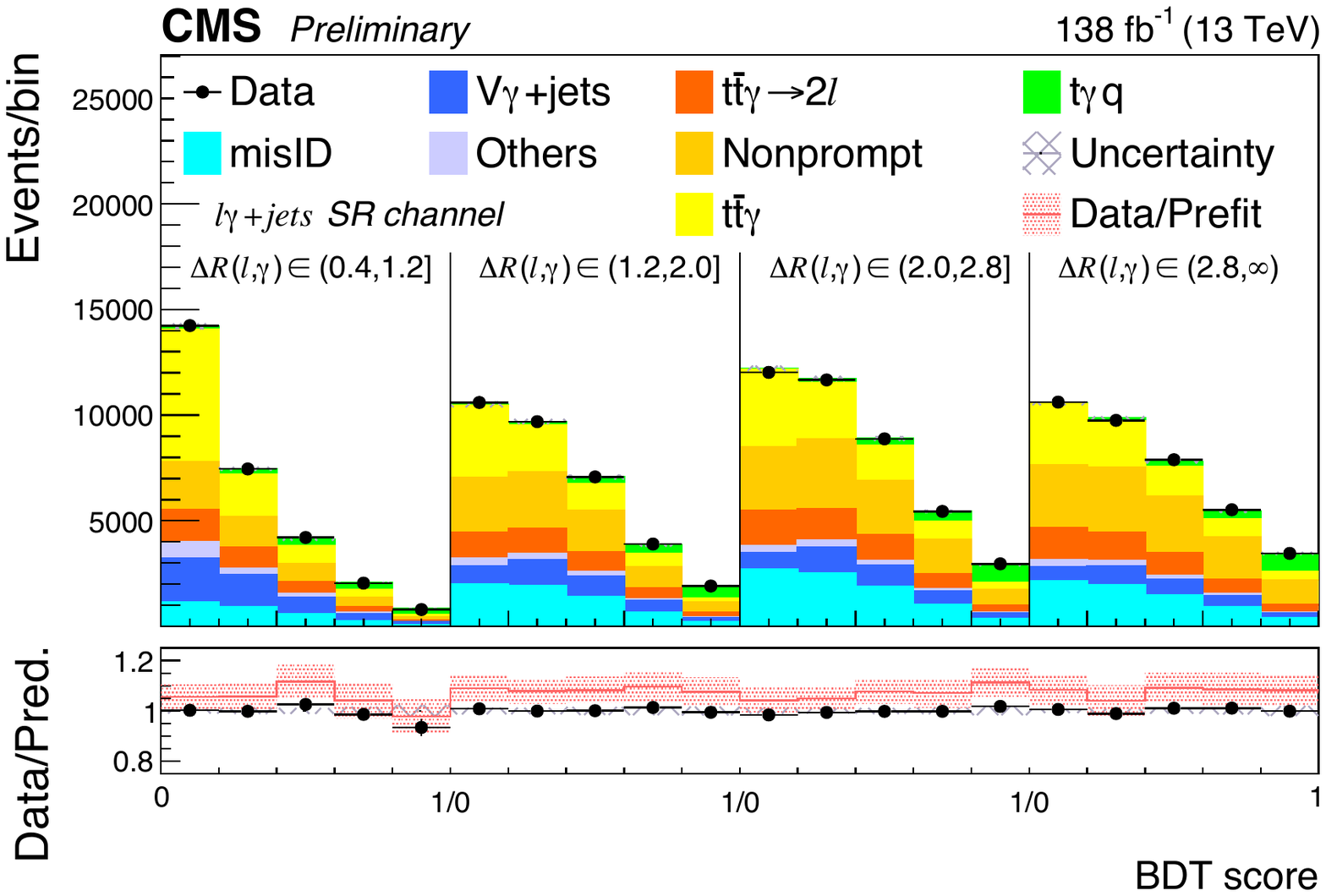}
  
  \includegraphics[width=0.45\textwidth]{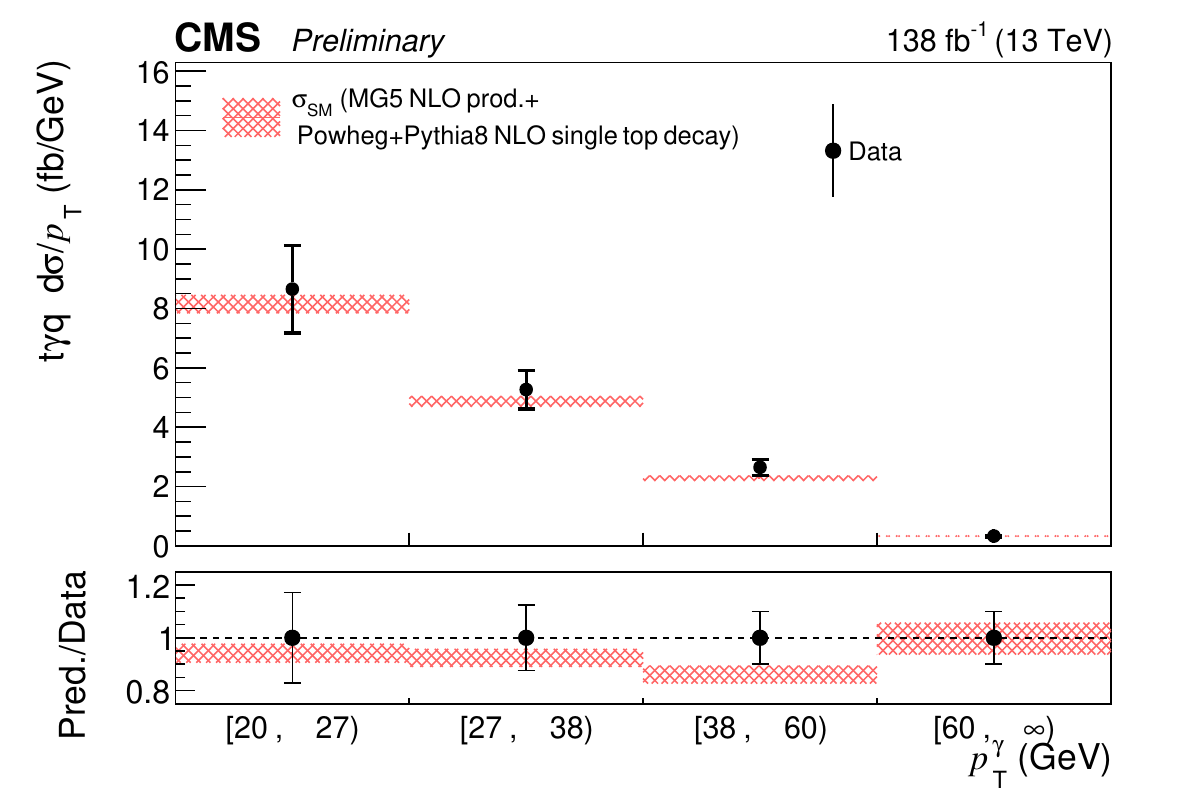}
  \includegraphics[width=0.45\textwidth]{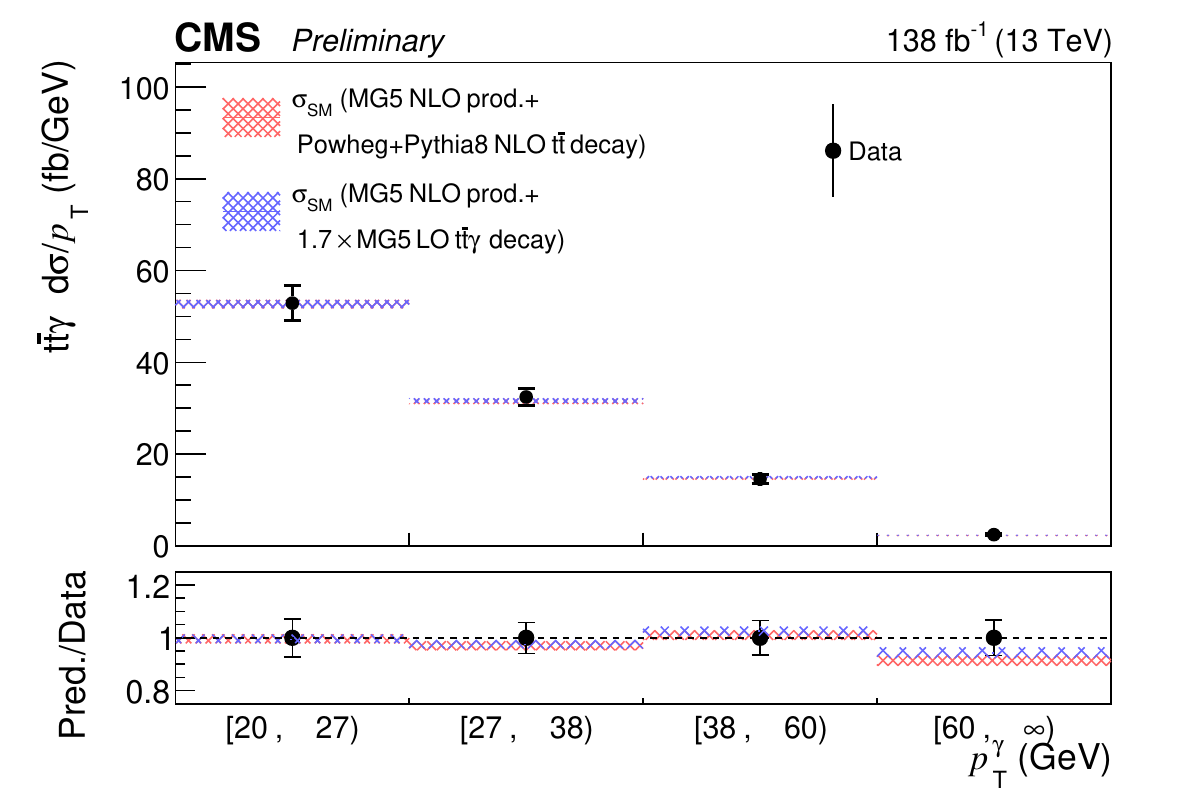}
  \caption{Upper: unrolled joint distribution of the BDT discriminant and $\Delta R(\ell,\gamma)$.
    Lower: unfolded differential cross section as a function of the photon \pt for \tqg production (left) and
    \ttg production (right). All plots are taken from Ref.~\protect\cite{tqg}.
  }
  \label{fig:tqgamma}
\end{figure}

A measurement of \ttg and \tqg production is reported by the CMS
Collaboration~\cite{tqg}. This result complements the first observation of \tqg
production by the ATLAS Collaboration~\cite{tqg_atlas} and provides the first
differential measurement of this process, performed simultaneously for \ttg and \tqg.
It also constitutes the first observation of \tqg production by the CMS Collaboration.

The analysis selects events with a single lepton, an isolated photon, and additional
hadronic activity, targeting the decay products of the top quark and the spectator
quark characteristic of \tqg production. A boosted decision tree (BDT) discriminant is
used to separate the \ttg and \tqg processes. Dedicated control regions are employed to
constrain the dominant backgrounds, which include events with electrons or hadronic jets
misidentified as photons, diboson production, and \ttg production in dilepton final
states.

The inclusive cross sections for \ttg and \tqg production are extracted using a
simultaneous fit to a two-dimensional distribution defined by the BDT output and the
angular separation between the lepton and the photon, $\Delta R(\ell,\gamma)$, shown in
Fig.~\ref{fig:tqgamma}. The measurement of the two signal processes is performed
simultaneously in all cases. The measured cross sections are $1445 \pm 80$~fb and $236 \pm 17$~fb for \ttg and \tqg production, respectively, in agreement with the SM
predictions of $1369 \pm 23$~fb and $207 \pm 9$~fb. Differential cross sections are also
measured as a function of several kinematic variables, all showing good agreement with
SM predictions. As an example, the differential cross section as a function of the photon
\pt is shown in Fig.~\ref{fig:tqgamma}.

\section{Observation of associated production of top quarks with boson pairs}

The ATLAS and CMS Collaborations have reported the observation of \ttgg
production~\cite{ttgg} and \tWZ production~\cite{twz}, respectively. These processes
probe rare electroweak interactions of the top quark and provide sensitivity to possible
anomalous couplings.

\begin{figure}[h]
  \centering
  \includegraphics[width=0.4\textwidth]{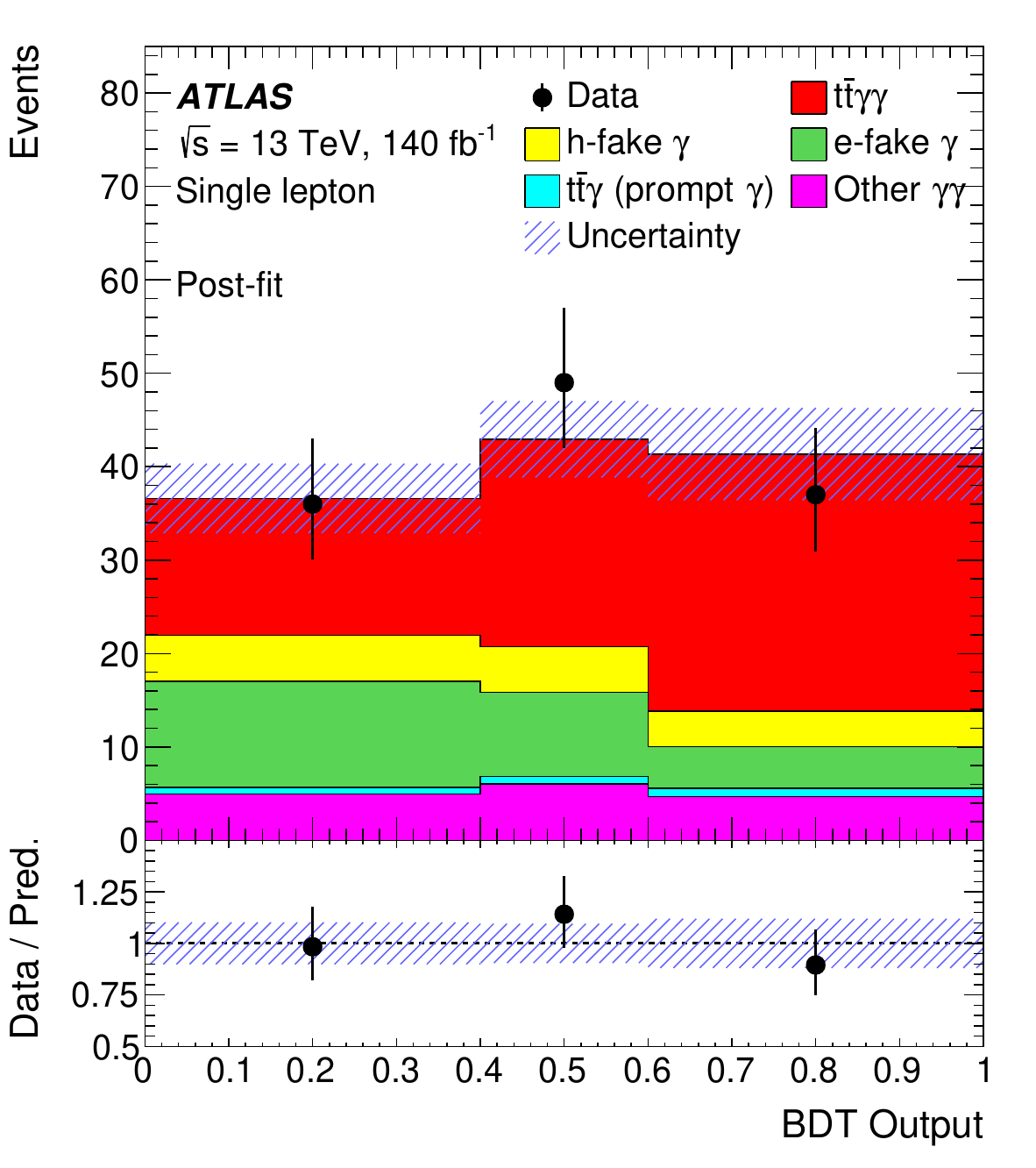}
  \includegraphics[width=0.4\textwidth]{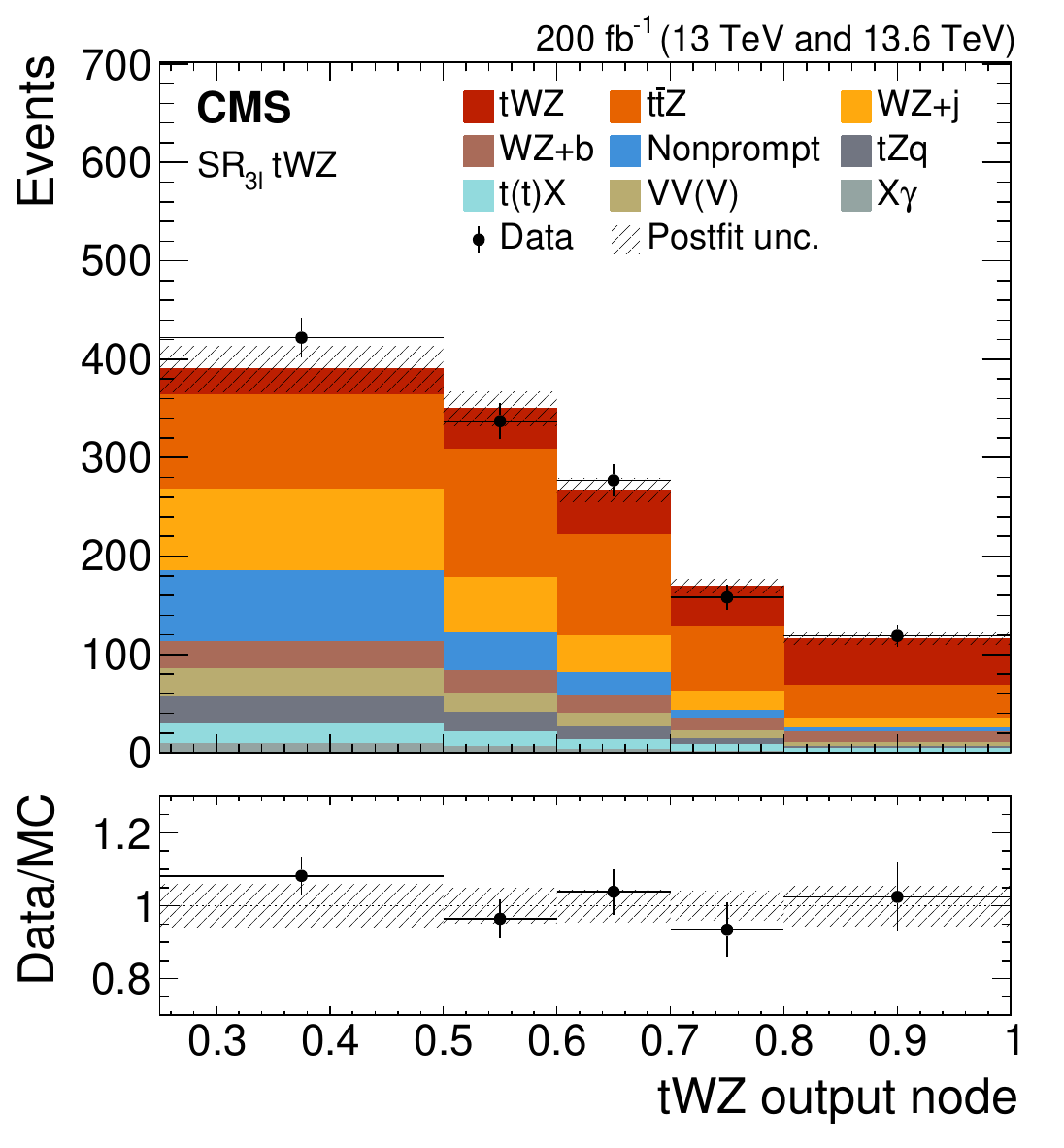}
  \includegraphics[width=0.44\textwidth]{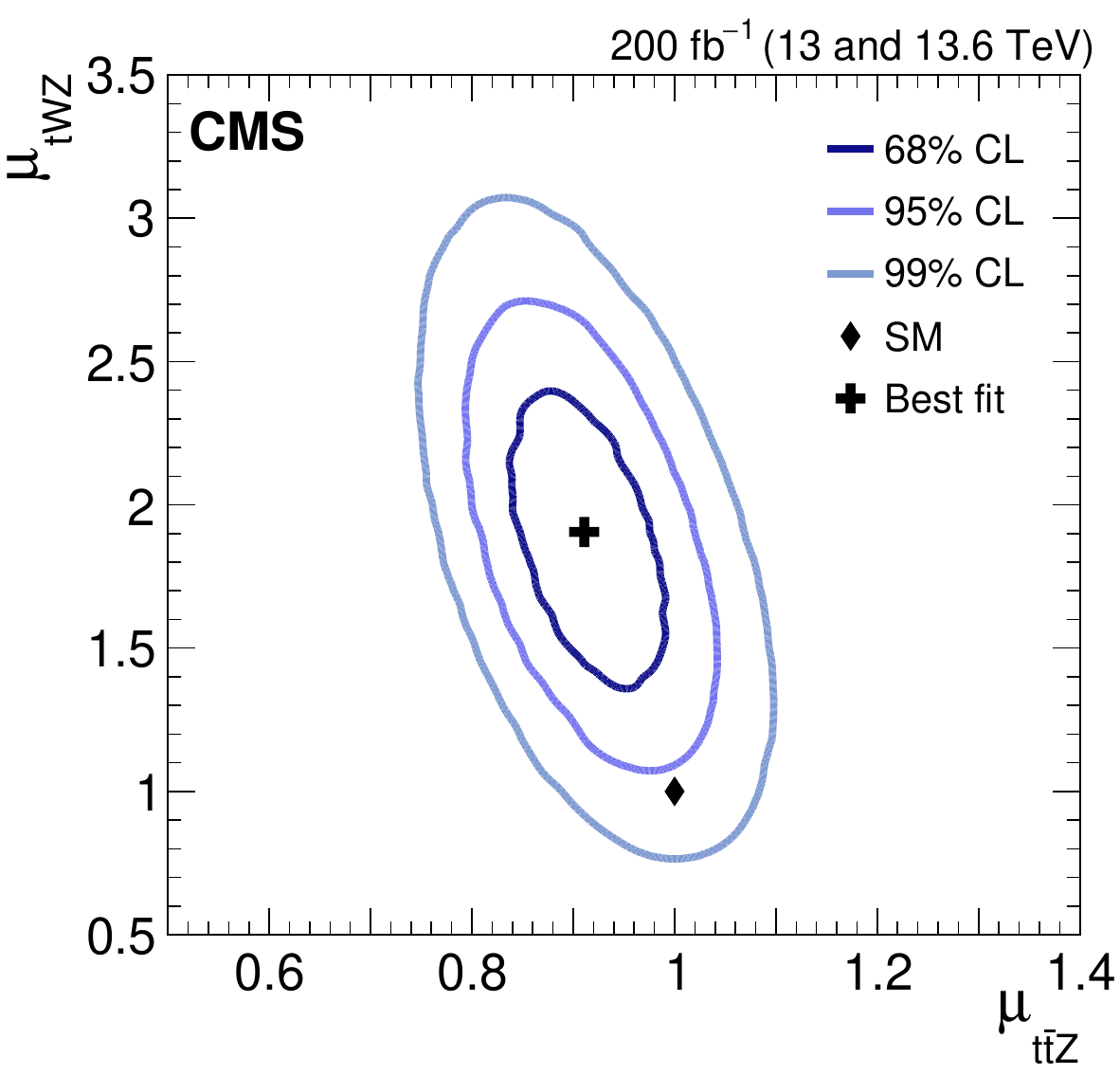}
  \caption{Upper left: distribution of the BDT score used in the observation of \ttgg production, taken from Ref.~
    \protect\cite{ttgg}. Upper right: distribution of the PartT output employed in the observation of \tWZ, taken from
  Ref.~\protect\cite{twz}. Lower: two-dimensional confidence regions for the simultaneous measurement of \ttZ and \tWZ production, taken from Ref.~\protect\cite{twz}.}
  \label{fig:top_diboson}
\end{figure}

The observation of \ttgg production uses events containing a single lepton, a pair of
photons, and additional hadronic activity. The event selection provides high purity in
\ttgg production, and a BDT discriminant, shown in Fig.~\ref{fig:top_diboson}, is employed to enhance the
sensitivity. The signal
is extracted using a likelihood fit to the BDT distribution, resulting in a measured
cross section of $2.42^{+0.58}_{-0.53}$~fb. The SM prediction for this process is currently
available at leading order in perturbative QCD, yielding $1.5^{+0.5}_{-0.4}$~fb, with an
expected $k$-factor of approximately 1.7~\cite{ttgg}. The measurement is therefore
consistent with theoretical expectations.

The observation of \tWZ production builds upon previous evidence reported by the CMS
Collaboration~\cite{twz_run2}. A major challenge in this measurement is the separation
of the signal from the irreducible \ttZ background, which exhibits a similar final-state
topology. This separation is improved through the use of a discriminator based on the
PartT architecture~\cite{part} shown in Fig.~\ref{fig:top_diboson}. The measurement combines data collected at
$\sqrt{s}=13$ and 13.6 TeV and results in an observed (expected) significance of 5.8
(3.5) standard deviations relative to the background-only hypothesis. A simultaneous
measurement of \ttZ and \tWZ production is also performed, demonstrating enhanced
separation between the two processes.

\section{Search for \texorpdfstring{\ttt}{ttt} production}

A search for \ttt production is presented in Ref.~\cite{ttt}. This process is highly
suppressed in the SM, as it proceeds only through electroweak interactions,
but it can be enhanced by new physics contributions generating flavor-changing neutral
currents. The analysis considers events with one lepton and events containing two or
more same-sign leptons. The latter category provides the highest sensitivity and is
therefore emphasized in this contribution.

\begin{figure}[h]
  \centering
  \includegraphics[width=0.4\textwidth]{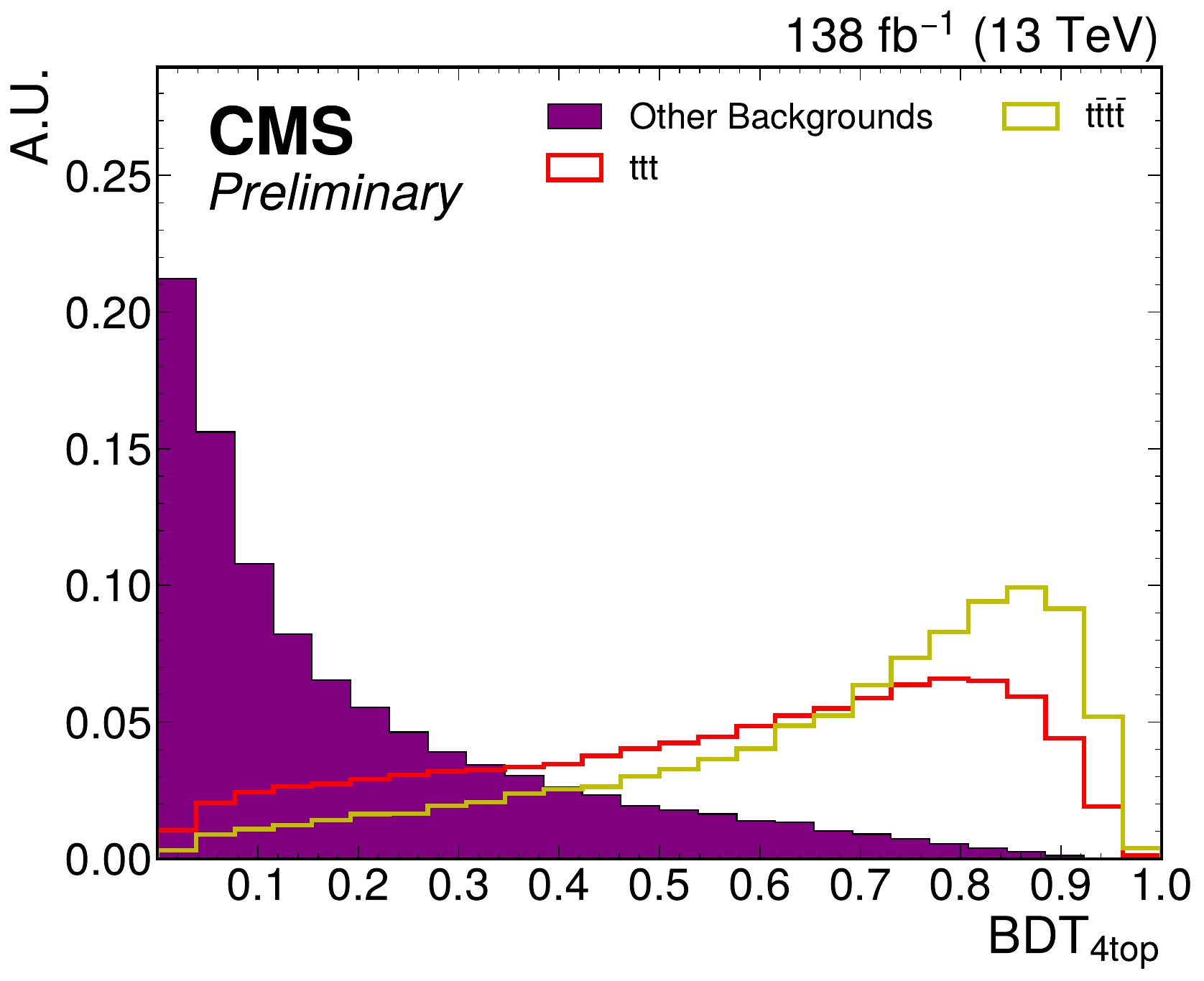}
  
  \includegraphics[width=0.4\textwidth]{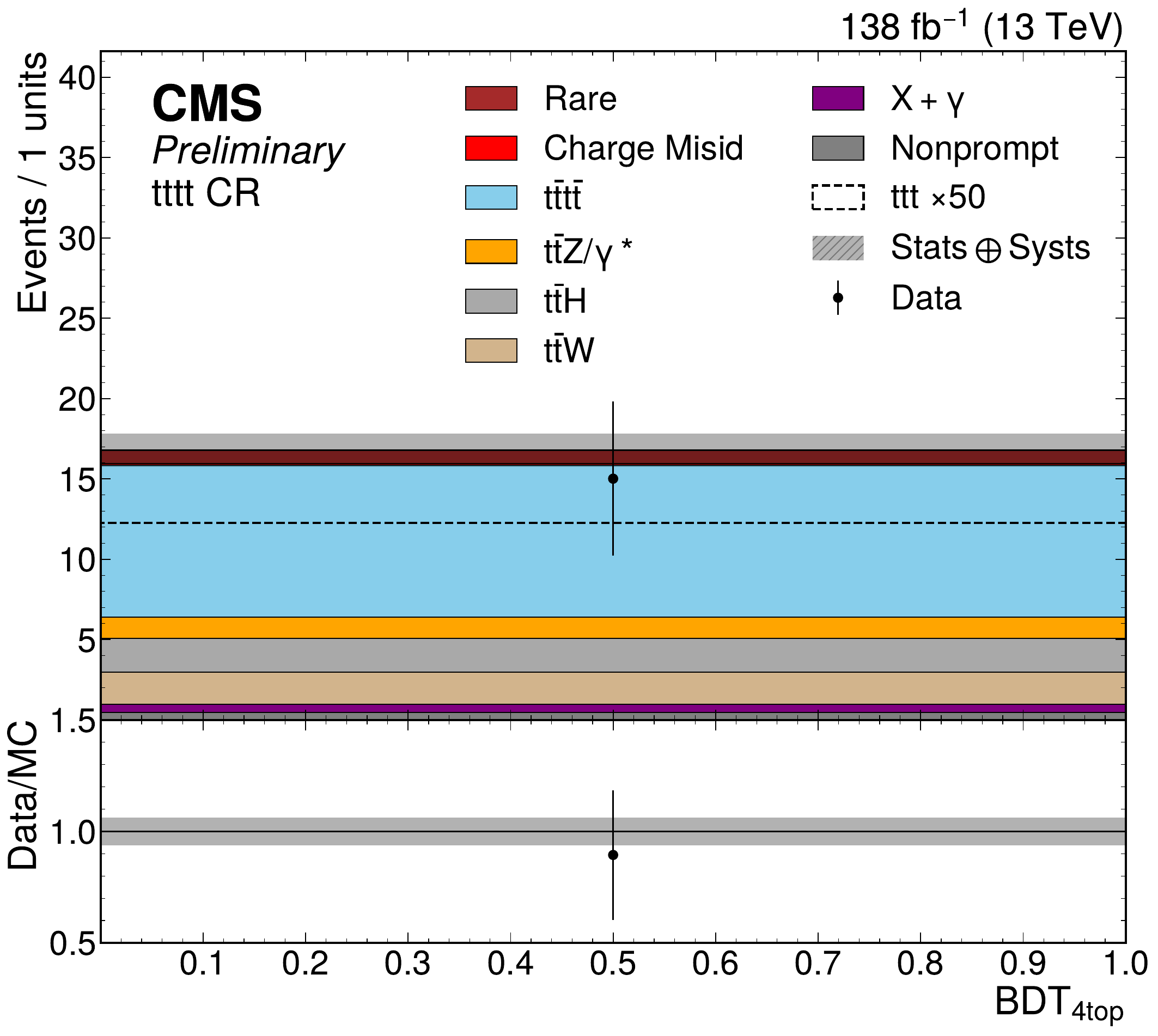}
  \includegraphics[width=0.4\textwidth]{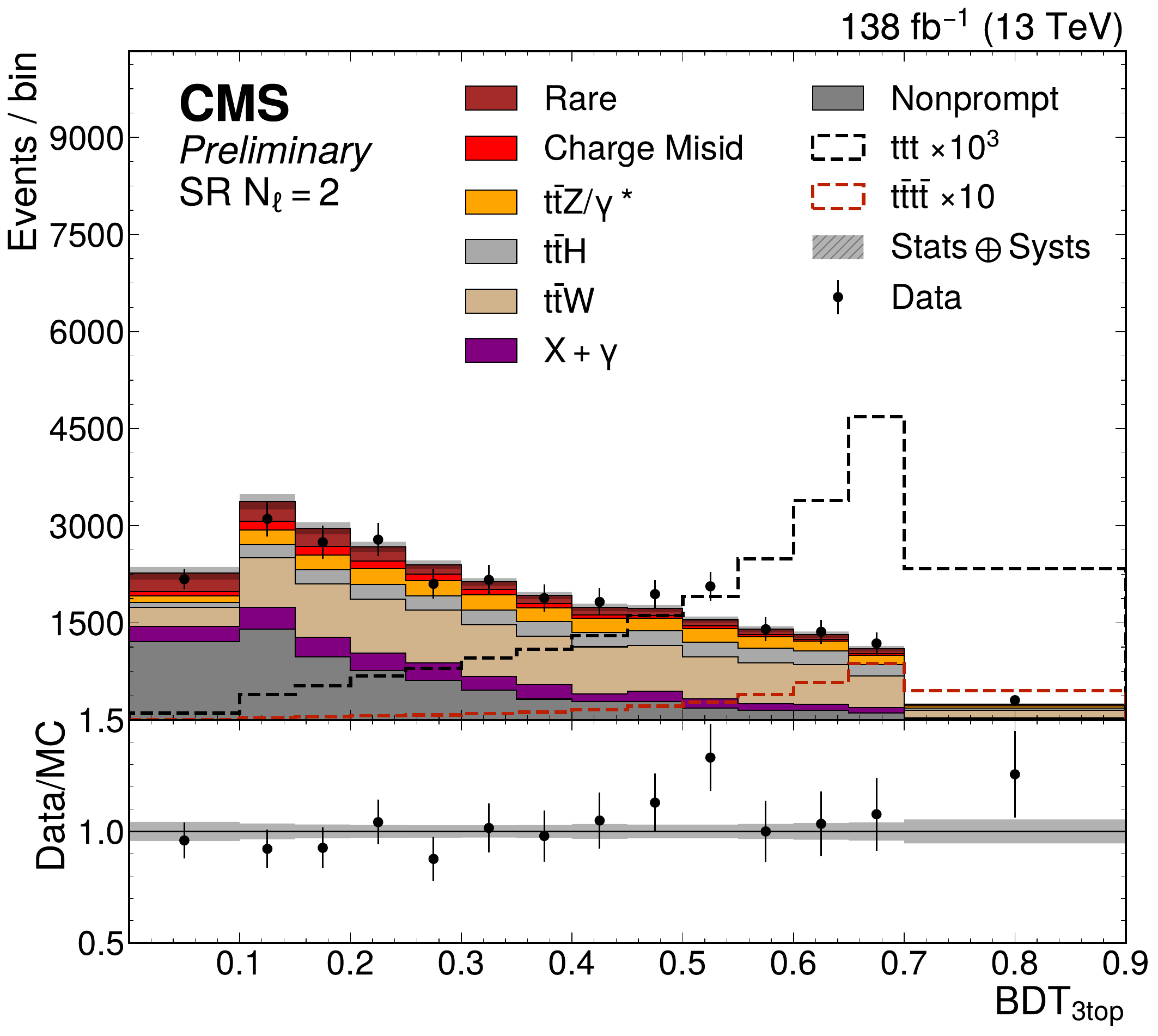}
  \caption{Upper: distribution of the BDT employed to discriminate backgrounds and \ttt production
    from \tttt production. Lower left: events in the \tttt control region.
    Lower right: distribution of the BDT employed to discriminate
  \ttt from the remaining backgrounds. All plots are taken from Ref.~\protect\cite{ttt}.}
  \label{fig:ttt}
\end{figure}

In addition to the lepton requirements, events must contain significant hadronic
activity. The dominant background contributions arise from \ttW production and events
with nonprompt leptons. The latter is estimated using a data-driven method, profiting from a
selection in which the lepton identification criteria are loosened to achieve a region
enriched in this background.
The contribution from \tttt production is particularly difficult
to disentangle from the signal. To address this, a BDT is trained to separate \tttt from
the signal and other backgrounds, allowing the construction of a control region enriched
in \tttt events, shown in Fig.~\ref{fig:ttt}. A second BDT discriminant, shown in Fig.~\ref{fig:ttt}, is then used to
separate \ttt production from the remaining backgrounds.

The analysis sets an observed (expected) upper limit of 25 (26)~fb on the \ttt cross
section, compared with a SM prediction of approximately 2~fb. The results
are consistent with expectations and set more stringent constraints in \ttt than previous
measurements.

\section*{Acknowledgments}

The author's work was supported by the ``Ram\'on y Cajal'' program under Project No.
RYC2024-048719-I, funded by ICIU/AEI/10.13039/501100011033 and by the FSE+.

\end{document}